\documentclass[a4paper,12pt]{article}

\usepackage[english]{babel}
\usepackage[utf8x]{inputenc}
\usepackage[T1]{fontenc}
\usepackage{authblk}
\usepackage{natbib}
\usepackage{here}
\usepackage{setspace}
\usepackage{multirow}
\usepackage{lmodern}
\usepackage{amsmath, amssymb}
\usepackage{bm}
\usepackage{pdflscape}
\usepackage{authblk}
\usepackage{color}
\usepackage[table]{xcolor}
\usepackage{booktabs}
\usepackage{afterpage}
\usepackage{textcomp}
\usepackage{float}
\usepackage{enumerate}
\usepackage{rotating}

\definecolor{babyblue}{rgb}{0.54, 0.81, 0.94}
\definecolor{babypink}{rgb}{0.96, 0.76, 0.76}

\usepackage[margin=1in]{geometry}

\usepackage{amsmath}
\usepackage{graphicx}
\usepackage[colorlinks=true, allcolors=blue]{hyperref}

\usepackage{caption}
\usepackage{subcaption}

\newcommand{\pg}{P{\'o}lya-Gamma }

\definecolor{Gray}{gray}{0.85}
\newcolumntype{a}{>{\columncolor{Gray}}c}
\newcolumntype{d}{>{\columncolor{white}}c}

\begin{document}

    \begin{center}
        \vspace*{1cm}
        \large
	    \textbf{Joint species distribution models with imperfect detection for high-dimensional spatial data}\\
         \normalsize
           \vspace{5mm}
	    Jeffrey W. Doser\textsuperscript{1, 2}, Andrew O. Finley\textsuperscript{2, 3}, Sudipto Banerjee\textsuperscript{4}
         \vspace{5mm}
    \end{center}
    \small
	     \textsuperscript{1}Department of Integrative Biology, Michigan State University, East Lansing, MI, USA \\
         \textsuperscript{2}Ecology, Evolution, and Behavior Program, Michigan State University, East Lansing, MI, USA \\
         \textsuperscript{3}Department of Forestry, Michigan State University, East Lansing, MI, USA \\
         \textsuperscript{4}Department of Biostatistics, University of California, Los Angeles, CA \\
         \noindent \textbf{Corresponding Author}: Jeffrey W. Doser, email: doserjef@msu.edu; ORCID ID: 0000-0002-8950-9895 \\
         \noindent \textbf{Running Title}: Spatial JSDMs with imperfect detection

\section*{Open Research}

The package \texttt{spOccupancy} is available on the Comprehensive R Archive Network (CRAN; \url{https://cran.r-project.org/web/packages/spOccupancy/index.html}). Data and code used in the manuscript are available on GitHub (\url{https://github.com/doserjef/Doser_et_al_2022}) and will be posted on Zenodo upon acceptance. 
         
\newpage
\section*{Abstract}

Determining spatial distributions of species and communities are key objectives of ecology and conservation. Joint species distribution models use multi-species detection-nondetection data to estimate species and community distributions. The analysis of such data is complicated by residual correlations between species, imperfect detection, and spatial autocorrelation. While methods exist to accommodate each of these complexities, there are few examples in the literature that address and explore all three complexities simultaneously. Here we developed a spatial factor multi-species occupancy model to explicitly account for species correlations, imperfect detection, and spatial autocorrelation. The proposed model uses a spatial factor dimension reduction approach and Nearest Neighbor Gaussian Processes to ensure computational efficiency for data sets with both a large number of species (e.g., > 100) and spatial locations (e.g., 100,000). We compare the proposed model performance to five candidate models, each addressing a subset of the three complexities. We implemented the proposed and competing models in the \texttt{spOccupancy} software, designed to facilitate application via an accessible, well-documented, and open-source R package. Using simulations, we found ignoring the three complexities when present leads to inferior model predictive performance, and the impacts of failing to account for one or more complexities will depend on the objectives of a given study. Using a case study on 98 bird species across the continental US, the spatial factor multi-species occupancy model had the highest predictive performance among the candidate models. Our proposed framework, together with its implementation in \texttt{spOccupancy}, serves as a user-friendly tool to understand spatial variation in species distributions and biodiversity while addressing common complexities in multi-species detection-nondetection data.     

%Further, our model successfully distinguished between two biogeographical species groups within the 98 species, indicating the potential of our framework as a model-based ordination technique.

\noindent \textbf{Keywords}: Bayesian, latent factor, Nearest Neighbor Gaussian Process, occupancy model 

\section*{Introduction}

Understanding the spatial distributions of species and communities is a fundamental task of ecology and conservation. Species distribution models (SDMs) are popular for predicting species distributions and their drivers across space and time \citep{guisan2000predictive}, which have informed key developments in ecological theory as well as conservation and management decisions \citep{bateman2020north}. While SDMs can use different data types, they most commonly use binary detection-nondetection data. Advances in hierarchical modeling have addressed many issues encountered when modeling multi-species detection-nondetection data. In particular, the three major complexities are (1) residual species correlations \citep{ovaskainen2010modeling}, (2) imperfect detection \citep{mackenzie2002}, and (3) spatial autocorrelation \citep{latimer2009hierarchical, banerjee2014hierarchical}.  

Joint species distribution models (JSDMs) are regression-based approaches that explicitly accommodate residual species correlations \citep{latimer2009hierarchical, ovaskainen2010modeling}. By jointly modeling species within a single model, JSDMs facilitate co-occurrence hypothesis testing \citep{ovaskainen2010modeling} and increase precision of both individual species distributions and community metrics. However, JSDMs typically do not accommodate imperfect detection (but see \citealt{tobler2019joint, hogg2021effectiveness}). Failure to account for imperfect detection in detection-nondetection data can lead to biases in both species distributions and the effects of environmental drivers on species occurrence \citep{mackenzie2002}. Occupancy models, a specific type of SDM, explicitly account for imperfect detection separately from the true species occurrence process using replicated detection-nondetection data. Multi-species occupancy models are an extension to single-species occupancy models that use detection-nondetection data from multiple species by treating species as random effects arising from a community-level distribution \citep{dorazio2005, gelfand2005modelling}. Unlike JSDMs, multi-species occupancy models do not estimate residual co-occurrence associations between species (but see \citealt{tobler2019joint}). 

Accounting for spatial autocorrelation in SDMs is often necessary when modeling species distributions across large spatial extents or a large number of observed locations \citep{latimer2009hierarchical}. Spatially-explicit SDMs account for spatial autocorrelation by including spatially-structured random effects \citep{banerjee2014hierarchical, shirota2019sinica}.  Such spatially-explicit approaches have been used in JSDMs to simultaneously account for residual species correlations and spatial autocorrelation \citep{thorson2015spatial}, and in multi-species occupancy models that model imperfect detection \citep{doser2022spoccupancy}. 

Despite development of JSDMs, multi-species occupancy models, and their spatially-explicit extensions, only recently have approaches emerged that incorporate species correlations and imperfect detection in SDMs for large communities \citep{tobler2019joint, hogg2021effectiveness}. Further, these approaches can become computationally intensive as both the number of spatial locations and species in the community increases, and no approaches exist that simultaneously incorporate species correlations, imperfect detection, and spatial autocorrelation, despite the well-recognized impacts of ignoring these complexities. Here we develop a joint species distribution model that explicitly accounts for species correlations, imperfect detection, and spatial autocorrelation. Analogous to \cite{tikhonov2020computationally}, we build an ecological process model that uses a spatial factor model together with Nearest Neighbor Gaussian Processes (NNGPs; \citealt{datta2016hierarchical}) to ensure computational efficiency for large species assemblages (e.g., > 100 species) across a large number of spatial locations (e.g., ${\sim} 10^5$). We extend the model of \cite{tikhonov2020computationally} by incorporating an observation sub-model that separately models imperfect detection from the latent ecological process. We use simulations and a case study on 98 bird species across the continental US to compare performance of our proposed model with five alternative models that fail to address all three complexities. Our proposed modeling framework, and its user-friendly implementation in the \texttt{spOccupancy} R package \citep{doser2022spoccupancy}, provides a computationally efficient approach that explicitly accounts for imperfect detection to deliver inference on individual species distributions, species co-occurrence patterns, and overall biodiversity metrics.  

\section*{Modeling Framework}

\subsection*{Process Model}

Let $\bm{s}_j$ denote the spatial coordinates of site $j$, for all $j = 1, \dots, J$ sites. Define $z_i(\bm{s}_j)$ as the true latent presence (1) or absence (0) of species $i$ at site $j$ for $i = 1, \dots, N$ species. We assume $z_i(\bm{s}_j)$ arises from a Bernoulli distribution following
\begin{equation}\label{zBern}
  z_i(\bm{s}_j) \sim \text{Bernoulli}(\psi_i(\bm{s}_j))\;,        
\end{equation}
where $\psi_i(\bm{s}_j)$ is the probability of occurrence for species $i$ at site $j$. We model $\psi_i(\bm{s}_j)$ as
\begin{equation}\label{psi}
  \text{logit}(\psi_i(\bm{s}_j)) = (\beta_{i, 1} + \text{w}^*_i(\bm{s}_j)) + \sum_{t = 2}^{p_{\psi}}x_t(\bm{s}_j)\beta_{i, t}\;,
\end{equation}
where $x_t(\bm{s}_j)$, for each $t = 2, \dots, p_\psi$, is an environmental covariate at site $j$, $\beta_{i, t}$ is a regression coefficient corresponding to $x_t(\bm{s}_j)$ for species $i$, $\beta_{i, 1}$ is the species-specific intercept, and $\text{w}^*_{i}(\bm{s}_j)$ is a species-specific latent spatial process. While not shown in Equation \ref{psi}, we can also include unstructured random intercepts that may affect species-specific occurrence probability. We seek to jointly model the species-specific spatial processes to account for residual correlations between species. For a small number of species (e.g., < 10), such a process can be estimated via a linear model of coregionalization framework \citep{gelfand2004, latimer2009hierarchical}. However, when the number of species is even moderately large (e.g., > 10), estimating such a joint process becomes computationally intractable. A viable solution to this problem is to use a spatial factor model \citep{hogan2004bayesian, ren2013hierarchical}, a dimension reduction approach that can account for correlations among a large number of species. Specifically, we decompose $\text{w}^*_i(\bm{s}_j)$ into a linear combination of $q$ latent variables (i.e., factors) and their associated species-specific coefficients (i.e., factor loadings). In particular, we have 
\begin{equation}\label{wstar}
  \text{w}^*_i(\bm{s}_j) = \bm{\lambda}_i^\top\textbf{w}(\bm{s}_j), 
\end{equation}
where $\bm{\lambda}_i^{\top}$ is the $i$th row of factor loadings from an $N \times q$ loading matrix $\bm{\Lambda}$, and $\textbf{w}(\bm{s}_j)$ is a $q \times 1$ vector of independent spatial factors at site $j$. We achieve computational improvements and dimension reduction by setting $q << N$, where often a small number of factors (e.g., $q = 5$) is sufficient \citep{taylor2019spatial, zhang2021biom}. We account for residual species correlations via their individual responses (i.e., loadings) to the $q$ latent spatial factors. Given a single factor, if two species commonly occur together beyond that which is explained by the covariates included in the model, the species-specific factor loadings will show positive correlation, whereas if one species tends to occur at locations where the other is not present, the species-specific factor loadings will show negative correlation. The residual inter-species covariance matrix $\bm{\Sigma} = \bm{\Lambda}\bm{\Lambda}^{\top}$ has rank $q << N$ and, hence, is singular. \cite{shirota2019sinica} discuss its use and interpretation in detecting species clustering.

Following \cite{taylor2019spatial} and \cite{tikhonov2020computationally}, we model $\text{w}_r(\bm{s}_j)$ using an NNGP \citep{datta2016hierarchical} for each $r = 1, \dots, q$ to achieve computational efficiency when modeling a large number of spatial locations. More specifically, we have

\begin{equation} \label{spatialProcess}
    \text{w}_r(\bm{s}_j) \sim N(\bm{0}, \tilde{\bm{C}}_r(\bm{\theta}_r)),
\end{equation}
where $\tilde{\bm{C}}_r(\bm{\theta}_r)$ is the NNGP-derived covariance matrix for the $r^{\text{th}}$ spatial process. The vector $\bm{\theta}_r$ consists of parameters governing the spatial process according to a spatial correlation function \citep{banerjee2014hierarchical}. For many correlation functions (e.g., exponential, spherical, Gaussian), $\bm{\theta}_r$ includes a spatial variance parameter, $\sigma^2_r$, and a spatial range parameter, $\phi_r$, while the Mat\'ern correlation function includes an additional spatial smoothness parameter, $\nu_r$. 

We assume all species-specific parameters ($\beta_{i, t}$ for all $t = 1, \dots, p_{\psi}$) arise from community-level distributions to enable information sharing across species \citep{dorazio2005, gelfand2005modelling}. Specifically, we assign a normal prior with mean and variance hyperparameters that represent the community-level average and variance among species-specific effects across the community, respectively. For example, we model the species-specific occurrence intercept, $\beta_{i, 1}$, following
\begin{equation} \label{commEffects}
  \beta_{i, 1} \sim N(\mu_{\beta_1}, \tau^2_{\beta_1}), 
\end{equation} 
where $\mu_{\beta_1}$ and $\tau^2_{\beta_1}$ are the community-level average and variance, respectively.

\subsection*{Observation Model}

To estimate $\psi_i(\bm{s}_j)$ and $z_i(\bm{s}_j)$ while explicitly accounting for imperfect detection, we obtain $k = 1, \dots, K_j$ sampling replicates at each site $j$. Let $y_{i, k}(\bm{s}_j)$ denote the detection (1) or nondetection (0) of species $i$ during replicate $k$ at site $j$. We model the observed data $y_{i, k}(\bm{s}_j)$ conditional on the true species-specific occurrence $z_i(\bm{s}_j)$ at site $j$ following
\begin{equation}
    y_{i, k}(\bm{s}_j) \mid z_i(\bm{s}_j) \sim \text{Bernoulli}(\pi_{i, k}(\bm{s}_j)z_i(\bm{s}_j)),
\end{equation}
where $\pi_{i, k}(\bm{s}_j)$ is the probability of detecting species $i$ at site $j$ during replicate $k$ given the species is present at the site (i.e., $z_i(\bm{s}_j) = 1$). We model $\pi_{i, k}(\bm{s}_j)$ as a function of site and/or replicate-level covariates that may influence species-specific detection probability. Specifically, we have

\begin{equation}
    \text{logit}(\pi_{i, k}(\bm{s}_j)) =  \alpha_{i, 1} + \sum_{t = 2}^{p_{\pi}} v_{t, k}(\bm{s}_j) \alpha_{i, t},
\end{equation}
where $v_{t, k}(\bm{s}_{j})$ is the value of covariate $t$ at site $j$ during replicate $k$, $\alpha_{i, t}$ is a regression coefficient corresponding to $v_{t, k}(\bm{s}_j)$, and $\alpha_{i, 1}$ is a species-specific intercept. If applicable, we can also include unstructured random intercepts in the model for species-specific detection probability. Analogous to the species-specific occurrence effects (Equation \ref{commEffects}), we assume all species-specific detection parameters (i.e., $\alpha_{i, t}$ for all $t = 1, \dots, p_{\pi}$) arise from community-level normal distributions. 

\subsection*{Prior specification and identifiability considerations}

We assume normal priors for mean parameters and inverse-Gamma priors for variance parameters. Following \cite{taylor2019spatial}, we set all elements in the upper triangle of the factor loadings matrix $\bm{\Lambda}$ equal to 0 and its diagonal elements equal to 1 to ensure identifiability of the spatial factors. We additionally fix the spatial variance parameters $\sigma^2_{r}$ to 1. We assign standard normal priors for all lower triangular elements in $\bm{\Lambda}$ and assign each spatial range parameter $\phi_{r}$ an independent uniform prior.

\subsection*{Model implementation and candidate models}

We implement the spatial factor multi-species occupancy model in a Bayesian framework in the function \texttt{sfMsPGOcc} within our open-source \texttt{spOccupancy} R package \citep{doser2022spoccupancy}. We employ the computational considerations discussed in \cite{finley2020spnngp} to ensure spatially-explicit models are computationally feasible for large data sets. The Bayesian framework allows us to easily calculate biodiversity metrics, with fully propagated uncertainty, as derived quantities. For example, we can estimate species richness of the entire community (or a subset of species in the community) by summing up the latent occurrence state $z_{i}(\bm{s}_j)$ at each site $j$ for all species of interest at each iteration to yield a full posterior distribution for species richness. We leverage a \pg data augmentation scheme \citep{polson2013} to yield an efficient Gibbs sampler (see Appendix S2 for full details). 

We compare the spatial factor multi-species occupancy model to five candidate models that only address a subset of the three complexities (Appendix S1: Table S1). We provide functionality for all five candidate models in the \texttt{spOccupancy} R package, and subsequently refer to all models by their \texttt{spOccupancy} function name (Appendix S1: Table S1). Our first candidate model is a non-spatial latent factor JSDM (\texttt{lfJSDM}) that does not account for imperfect detection, analogous to standard JSDM approaches \citep{wilkinson2019comparison}. Our second candidate model is a spatial factor JSDM (\texttt{sfJSDM}) that does not account for imperfect detection, similar to the NNGP model of \cite{tikhonov2020computationally}. Our third model is the basic non-spatial multi-species occupancy model (\texttt{msPGOcc}) that does not incorporate residual species correlations \citep{dorazio2005}. Our fourth model is a spatial multi-species occupancy model (\texttt{spMsPGOcc}) that does not incorporate residual species correlations and estimates a separate spatial process for each species \citep{doser2022spoccupancy}. Finally, our fifth model is a non-spatial latent factor multi-species occupancy model (\texttt{lfMsPGOcc}) that accounts for residual species correlations and imperfect detection, analogous to the model of \cite{tobler2019joint}, except we use a logit formulation of the model. See Appendices S1 and S2 for full model details. 

\section*{Simulation Study}

We used simulations to compare estimates from the spatial factor multi-species occupancy model to estimates from the five candidate models (Appendix S1: Table S1). We generated 100 detection-nondetection data sets for each of six simulation scenarios, where the data were simulated with different combinations of the three complexities. We simulated data under situations that roughly corresponded to the six candidate models to assess how each model performed under ``ideal'' data conditions for that model, as well as when the data do not meet all the assumptions of the modeling framework. More specifically, we generated data with (1) residual species correlations and constant imperfect detection, (2) residual species correlations, constant imperfect detection, and spatial autocorrelation, (3) imperfect detection only, (4) imperfect detection and spatial autocorrelation, (5) residual species correlations and imperfect detection, and (6) residual species correlations, imperfect detection, and spatial autocorrelation. 

We simulated detection-nondetection data from $N = 10$ species at $J = 225$ sites with $K = 3$ replicates at each site for each of the 100 data sets for the six simulation scenarios. We used an exponential correlation function for spatially-explicit data generation scenarios (Scenarios 2, 4, 6). For scenarios leveraging a factor model (Scenarios 1, 2, 5, 6), we generated the data using $q = 3$ latent factors. As there are often many potential covariates that explain multi-species occurrence patterns in empirical data sets, we simulated data with 15 spatially-varying occurrence covariates for all scenarios and five observational-level detection covariates for scenarios where detection probability was not constant (Scenarios 3-6). We specified reasonable values for all parameters in the model (see Appendix S1 for full details). For each data set in each scenario, we ran three chains each of 15,000 samples, with a burn-in of 10,000 samples and a thinning rate of 5, resulting in a total of 3,000 MCMC samples for each of the six candidate models. We fit all models using the \texttt{spOccupancy} R package \citep{doser2022spoccupancy}. We assessed performance of the models by comparing the root mean squared error and 95\% coverage rates for the species-specific occurrence probabilities and the occurrence covariate effect.  

\section*{Case Study}

We applied the spatial factor multi-species occupancy model to detection-nondetection data from the North American Breeding Bird Survey \citep{pardieck2020north} in 2018 on $N = 98$ bird species at $J = 2619$ routes (i.e., sites) across the continental USA. The 98 species belong to two distinct biogeographical communities following the definitions in \cite{bateman2020north}, with 66 species in the eastern forest bird community and 32 species in the grassland bird community. Our objectives for this case study were to (1) develop spatially-explicit maps of species richness for the two communities across the continental USA, (2) determine if the latent spatial factors (\textbf{w}) and the species-specific factor loadings ($\bm{\Lambda}$) distinguish the two communities of birds, and (3) assess the benefits of accounting for species correlations, imperfect detection, and spatial autocorrelation. At 50 points along each route (called ``stops''), observers performed a three-minute point count survey of all birds seen or heard within a 0.4km radius. We summarized the data for each species at each site into $K = 5$ spatial replicates (each comprising data from 10 of the 50 stops), where each spatial replicate took value 1 if the species was detected at any of the 10 stops in that replicate, and value 0 if the species was not detected. 

Using the spatial factor multi-species occupancy model, we modeled route-level occurrence of the 98 species as a function of five bioclimatic variables and eight land cover variables (Appendix S1). We modeled detection as a function of the day of survey (linear and quadratic), time of day (linear), and a random observer effect. We standardized all variables to have a mean of 0 and standard deviation of 1. We fit the model using 15 nearest neighbors, an exponential correlation function, and $q = 5$ latent spatial factors. We subsequently predicted occurrence for the 98 species across the continental USA to generate spatially-explicit maps of species richness, with associated uncertainty, for the two bird communities.

To determine if the spatial factor multi-species occupancy model provided benefits for predicting species distributions and biodiversity metrics, we fit four additional candidate models (\texttt{msPGOcc}, \texttt{lfMsPGOcc}, \texttt{lfJSDM}, \texttt{sfJSDM}). For the models that do not explicitly model imperfect detection (\texttt{lfJSDM} and \texttt{sfJSDM}), we collapsed the data with five replicates at each site into a single binary value, which takes value 1 if the species was detected in any of the five replicates and 0 if not. Additionally, because the detection covariates we include in our model only vary by site and not by replicate, we included the detection covariates together with the occurrence covariates in the two JSDMs without a distinct submodel, which is a common approach used to account for sampling variability in models that do not explicitly account for imperfect detection \citep{ovaskainen2017make}. We used the Widely Applicable Information Criterion (WAIC; \citealt{watanabe2010}) to compare the performance of the three occupancy models (\texttt{msPGOcc}, \texttt{lfMsPGOcc}, and \texttt{sfMsPGOcc}) and the two JSDMs (\texttt{lfJSDM} and \texttt{sfJSDM}). However, since the two JSDMs use a collapsed form of the data used in the occupancy models, we cannot directly compare all five models using WAIC. Thus, we additionally fit all models using 75\% of the data points and kept the remaining 25\% of the data points for evaluation of model predictive performance. We assessed out-of-sample predictive performance using the observed data at the hold-out locations as well as latent occupancy predictions at the hold-out locations generated from models that used the entire data set. See Appendix S1: Section S3 for details. We ran all models in \texttt{spOccupancy} for three chains, each with 150,000 iterations with a burn-in period of 100,000 iterations and a thinning rate of 50.

\section*{Results}

\subsection*{Simulation study}

Failing to account for residual species correlations had negative impacts on both the accuracy and precision of model estimates (Tables \ref{tab:coverage}, Appendix S1: Tables S2, S3). Estimates from \texttt{msPGOcc}, which does not account for residual species correlations, had larger bias (Appendix S1: Tables S2, S3), and low coverage rates (Table \ref{tab:coverage}) for both latent occurrence and covariate effects when data were simulated with residual correlations between species. \texttt{spMsPGOcc}, which accounts for spatial autocorrelation but ignores species correlations, had less bias and better coverage rates than \texttt{msPGOcc} in these scenarios, but still had higher bias in occurrence probabilities and lower coverage rates than models that did account for species correlations. This suggests that accounting for spatial autocorrelation can mitigate some, but not all, of the negative impacts of incorrectly assuming independence between species.

When data were simulated with imperfect detection that varied across sites and replicates, ignoring imperfect detection resulted in higher bias and low coverage rates for both occurrence probability and covariate effects (Table \ref{tab:coverage}, Appendix S1: Tables S2, S3). However, when detection was high and constant over sites and replicates (Scenarios 1 and 2), bias in \texttt{lfJSDM} and \texttt{sfJSDM} was comparable to models that address imperfect detection and coverage rates were closer to the expected 95\%, in particular for the latent occurrence probability (Appendix S1: Tables S2, S3). Notably, the decreased coverage rates were less drastic for estimating occurrence probability when failing to account for imperfect detection compared to estimates from a standard multi-species occupancy model (\texttt{msPGOcc}) when ignoring residual correlations when present. Alternatively, failing to account for imperfect detection when present resulted in larger bias and smaller coverage rates in occurrence covariate effect estimates compared to a model that ignores residual correlations and/or spatial autocorrelation when present. Ignoring spatial autocorrelation had minimal impacts on average bias, but coverage rates were substantially low for both latent occurrence and the covariate effect for \texttt{msPGOcc} (Table \ref{tab:coverage}). 

\subsection*{Case study}

\begin{figure}[ht]
    \centering
    \includegraphics[width=15cm]{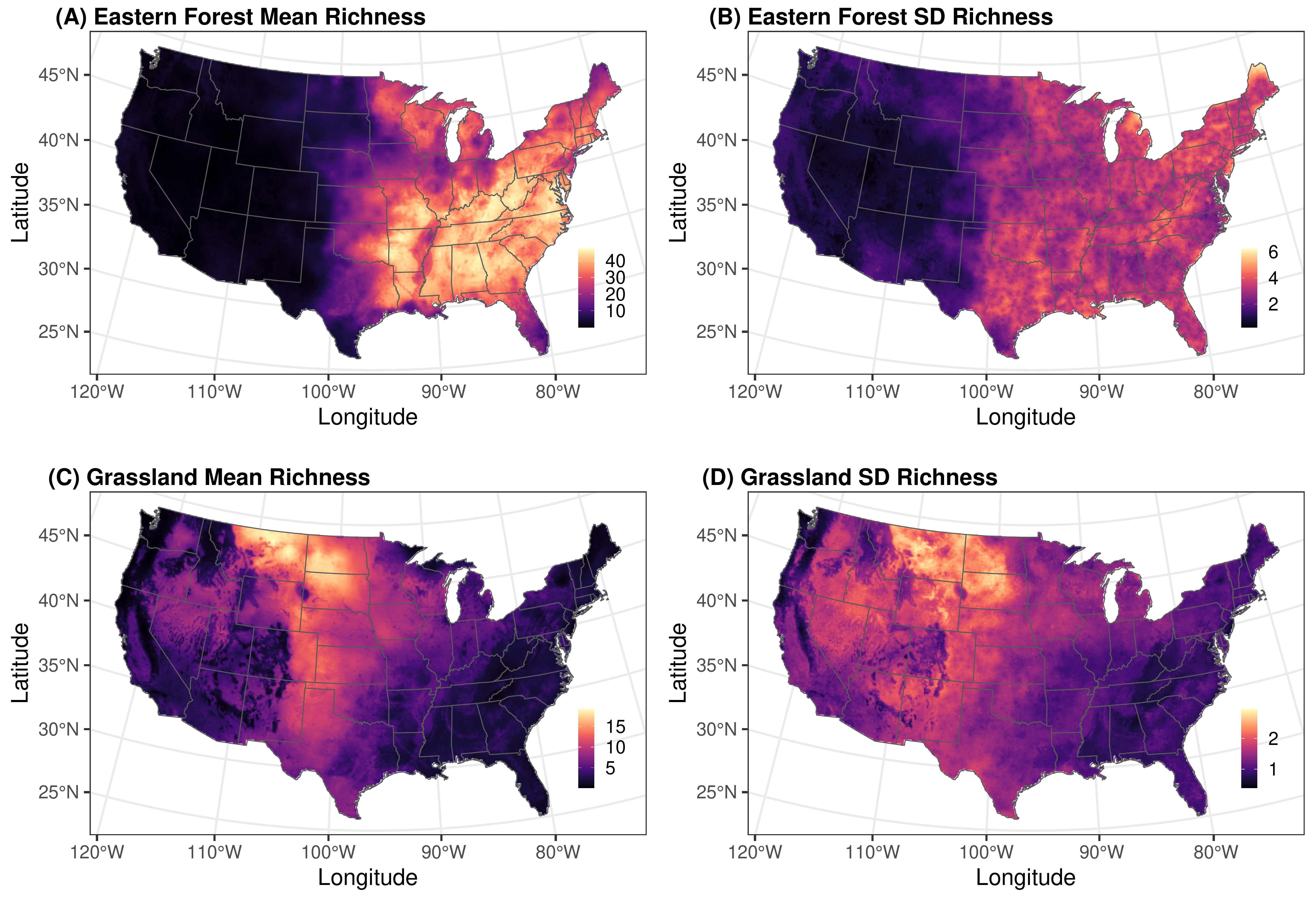}
    \caption{}
    \label{fig:richFigure}    
\end{figure}

\begin{figure}[ht]
    \centering
    \includegraphics[width=15cm]{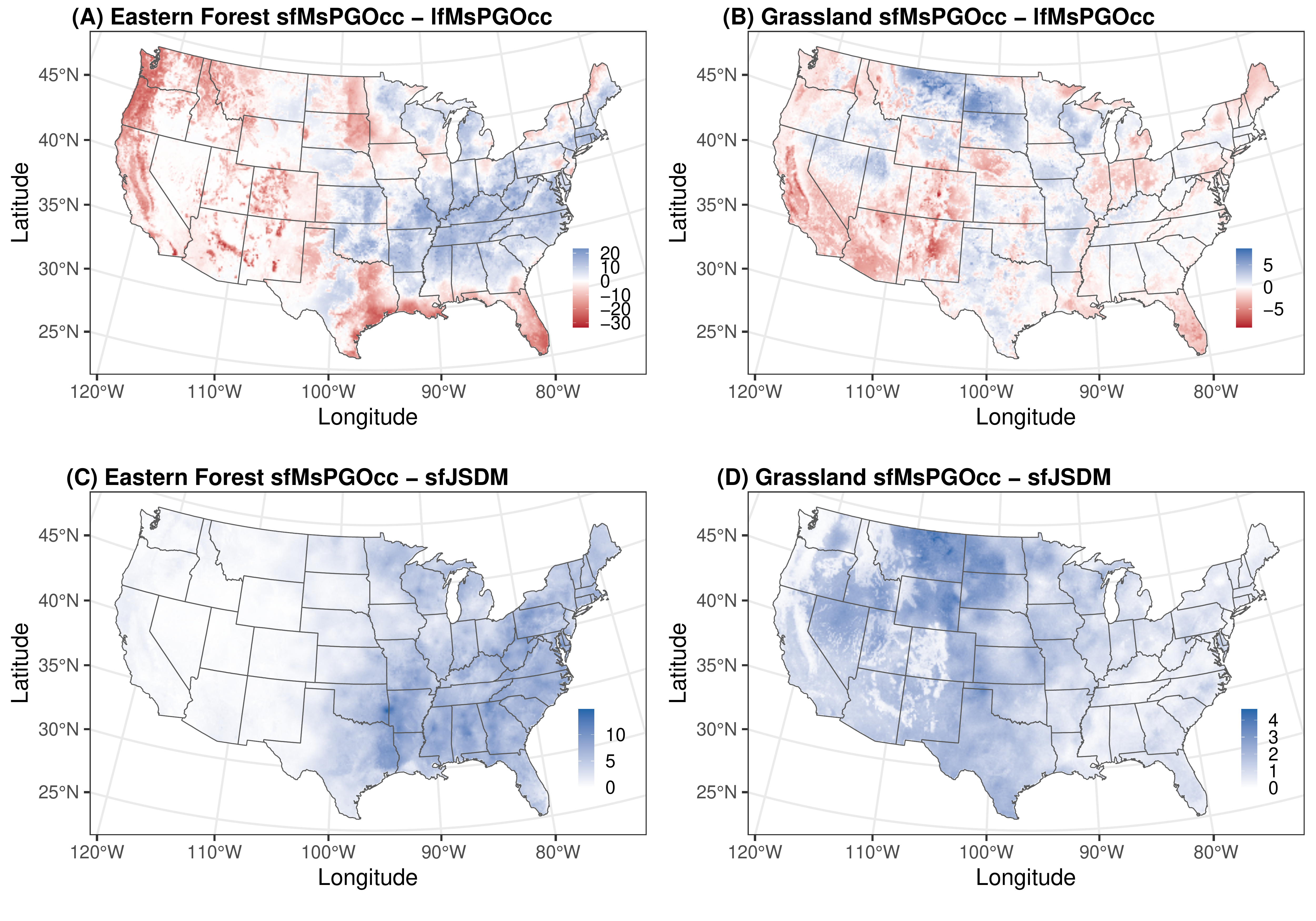}
    \caption{}
    \label{fig:richDiffFigure}    
\end{figure}

The spatial factor multi-species occupancy model predicted high species richness for the eastern forest bird community across the eastern US and high species richness for the grassland bird community in the Northern Great Plains region (Figure \ref{fig:richFigure}). Further, the model distinguished between the two bird communities via the species-specific factor loadings and the spatial factors (Appendix S1: Figures S1-S5). Compared to the standard multi-species occupancy model (\texttt{msPGOcc}), incorporating residual species correlations (\texttt{lfMsPGOcc}) yielded a lower WAIC (417,954 vs. 395,094), while additionally accounting for spatial autocorrelation (\texttt{sfMsPGOcc}) further reduced the WAIC (390,607; Appendix S1: Table S4). Failing to account for spatial autocorrelation led to unreasonable species richness estimates for the two communities across large portions of the US (Figure \ref{fig:richDiffFigure}A-B). Additionally, the spatially-explicit JSDM (\texttt{sfJSDM}) outperformed the non-spatial JSDM (\texttt{lfJSDM}) according to the WAIC (87,615 vs. 84,192). 

Analogous to model comparison using WAIC, the two models that accounted for spatial autocorrelation (\texttt{sfJSDM} and \texttt{sfMsPGOcc}) had the smallest out-of-sample model deviance, with \texttt{sfJSDM} outperforming \texttt{sfMsPGOcc} when assessing performance based on the raw detection-nondetection data. However, when estimating predictive performance using estimates of species occurrence generated from three occupancy model fits, \texttt{sfMsPGOcc} outperformed \texttt{sfJSDM} (Appendix S1: Table S4), suggesting that accounting for imperfect detection provides improved predictive performance of the latent ecological process. Further, estimates of species richness from \texttt{sfJSDM} were substantially lower across the US for both the eastern forest and grassland bird community (Figure \ref{fig:richDiffFigure}C-D) compared to estimates from \texttt{sfMsPGOcc}.

\section*{Discussion}

Multi-species detection-nondetection data are often complicated by residual correlations among species detections \citep{ovaskainen2010modeling}, imperfect detection of species \citep{mackenzie2002}, and spatial autocorrelation \citep{latimer2009hierarchical}. Here, we developed a spatial factor multi-species occupancy model that simultaneously accounts for all three complexities. We showed using simulations that ignoring these three complexities when present leads to inferior inference and prediction. Further, the spatial factor multi-species occupancy model improved predictive performance compared to models that failed to address the three complexities in an empirical case study of 98 bird species across the continental US. 

In our simulation study, failing to account for residual species correlations, imperfect detection, and/or spatial autocorrelation when present led to increased bias and low coverage rates. We found that the standard multi-species occupancy model (\texttt{msPGOcc}) had high bias and low coverage rates for both the latent occurrence and occurrence covariate effects for all scenarios except when data were simulated without species correlations and spatial autocorrelation (Table \ref{tab:coverage}, Appendix S1: Tables S2 and S3), clearly indicating the importance of accommodating these data complexities if they exist. Similarly, estimates from JSDMs that failed to account for imperfect detection resulted in increased bias and low coverage rates, although these findings were less prominent under ideal scenarios of constant, high detection probability. Interestingly, Table \ref{tab:coverage} suggests that if it is not possible to accommodate all three complexities (e.g., because of limited resources, small sample sizes) determining which complexities to ignore will depend on the study objectives. For example, when data were simulated with imperfect detection and species correlations, coverage rates were better for \texttt{lfJSDM} than \texttt{msPGOcc} for the occurrence probability estimates, but coverage rates from \texttt{msPGOcc} were better than \texttt{lfJSDM} for the occurrence covariate effect. This suggests that under these scenarios, \texttt{lfJSDM} would be better for prediction, while \texttt{msPGOcc} would be better for inference. While our simulation study did not consider all potential complexities when comparing the performance of occupancy models, these results do illustrate that specific data characteristics and research questions will determine whether it is necessary to account for residual species correlations, imperfect detection, and/or spatial autocorrelation. Our findings, as well as additional simulation studies geared towards specific ecological scenarios, could have important implications for designing detection-nondetection surveys to meet specific objectives. We include options to fit all six candidate models (Appendix S1; Table S1) in the \texttt{spOccupancy} R package, as well as functions for data simulation and model comparison to enable ecologists and conservation practitioners to accommodate these three complexities using accessible and well-documented software. See Appendix S3 for a detailed vignette on fitting these models in \texttt{spOccupancy} as well as the package website (\url{https://www.jeffdoser.com/files/spoccupancy-web/}) for additional tutorials. 

In the breeding bird case study, accounting for species correlations, imperfect detection, and spatial autocorrelation in the spatial factor multi-species occupancy model resulted in improved predictive performance compared to models that failed to address all three complexities. Accounting for species correlations in \texttt{lfMsPGOcc} improved model fit over the standard multi-species occupancy model (\texttt{msPGOcc}) according to WAIC but did not improve predictive performance for the out-of-sample deviance metric using the raw data (Appendix S1: Table S4). This is likely a result of treating the latent factors as independent standard normal random variables, which results in predictions that are not able to use the estimated values of the latent variables at nearby sampled locations to improve prediction at non-sampled locations. Alternatively, the spatial factor multi-species occupancy model (\texttt{sfMsPGOcc}) had the smallest WAIC and the best predictive performance for both deviance metrics among the three occupancy models. Further, \texttt{sfJSDM} substantially outperformed \texttt{lfJSDM} according to all criteria. These results demonstrate how assigning spatial structure to the latent factors in a model that accounts for species correlations can yield large improvements in model predictive performance. We thus recommend using \texttt{sfMsPGOcc} when there is a desire to account for species correlations and the primary goal of the analysis is prediction.

The spatial factor multi-species occupancy model leverages a spatial factor dimension reduction approach \citep{hogan2004bayesian, ren2013hierarchical} and NNGPs \citep{datta2016hierarchical} to ensure computational efficiency when modeling data sets with a large number of species (e.g., > 100) and/or spatial locations (e.g., 100,000). Our proposed model requires specification of the number of latent spatial factors ($q$) as well as the number of neighbors to use in the NNGP. When choosing the number of nearest neighbors for the NNGP, \cite{datta2016hierarchical} showed 15 neighbors is sufficient for most data sets, with as few as five neighbors providing adequate performance for certain data sets. Determining the optimal number of factors for a given data set is not straightforward and will vary depending on the characteristics of the specific community of species (e.g., species rarity, variability among species). See Appendix S4 for recommendations and considerations for making this decision.

The use of spatial replicates in the BBS case study instead of the more traditional temporal replicates used in an occupancy modeling framework may lead to upward bias in the estimated occupancy probabilities \citep{kendall2009cautionary}. Additionally, the large spatial scale of the BBS data (each route is $\sim$39.2km in length) likely influences the estimates of the residual species co-occurrence patterns. Data collected at a smaller spatial scale using temporal replicates may provide more accurate estimates of occupancy and species co-occurrence patterns. Regardless of how the data are collected, we caution against interpretation of the residual co-occurrences as true biological interactions, as co-occurrence does not imply an interaction \citep{poggiato2021interpretations}.

The latent spatial factors and the species-specific factor loadings can provide insight into the additional processes that govern distributions of species in the modeled community. In our case study, we found the spatial factors showed clear distinctions between the two bird communities. See Appendix S1 for additional discussion on interpreting the latent factors and Appendices S3 and S4 for practical information on how to troubleshoot MCMC convergence problems with the factor loadings. 

As both the number and size of multi-species detection-nondetection data sets increases, we require computationally efficient models and software to address common data complexities. Our spatial factor multi-species occupancy model extends previous approaches \citep{tobler2019joint, tikhonov2020computationally} to efficiently model species-specific and community-level occurrence patterns while accounting for residual species correlations, imperfect detection, and spatial autocorrelation. Our proposed framework, together with its user friendly implementation in the \texttt{spOccupancy} R package \citep{doser2022spoccupancy}, will enable ecologists to study spatial variation in species occurrence and co-occurrence patterns, develop spatially-explicit maps of individual species distributions and biodiversity metrics, and explicitly account for common complexities in multi-species detection-nondetection data.

\section*{Acknowledgements}

We thank Viviana Ruiz Gutierrez and an anonymous reviewer for insightful comments that improved the manuscript. This work was supported by National Science Foundation (NSF) grants EF-1253225 and DMS-1916395.

\bibliographystyle{apalike}
\bibliography{references}

\newpage 

\section*{Tables}

\begin{table}[ht!] % <--
  \begin{center}
  \caption{Estimated coverage rates of simulated species-specific occurrence probabilities and covariate effects for six different simulation scenarios and six models of varying complexity, as well as average run time. Coverage rates are defined as the percentage of species-specific occurrence probabilities ($\psi_i(\bm{s}_j)$) or covariate effects contained within the 95\% credible interval, averaged across the 10 species and 100 simulated data sets. Run time is the number of minutes for the model to complete 15,000 MCMC iterations, averaged across all six simulation scenarios and 100 simulated data sets.}
  \label{tab:coverage}
  \begin{tabular}{c c c c c c c c}
    \toprule
    Parameter & Scenario & & & Model & & & \\
    \midrule
     & & \texttt{lfJSDM} & \texttt{sfJSDM} & \texttt{msPGOcc} & \texttt{spMsPGOcc} & \texttt{lfMsPGOcc} & \texttt{sfMsPGOcc} \\
     $\psi_i(\bm{s}_j)$ & 1 & 91.5 & 90.8 & 68.9 & 88.1 & 95.6 & 95.3 \\
     & 2 & 91.6 & 91.0 & 69.1 & 89.1 & 95.5 & 95.4 \\
     & 3 & 85.6 & 84.8 & 95.0 & 96.4 & 95.5 & 95.5 \\
     & 4 & 77.5 & 76.4 & 80.2 & 93.1 & 95.7 & 95.5 \\
     & 5 & 75.3 & 74.2 & 71.3 & 88.5 & 95.5 & 95.3 \\
     & 6 & 76.0 & 75.0 & 72.2 & 89.6 & 95.3 & 95.2 \\
     $\beta_{i}$ & 1 & 88.7 & 88.2 & 82.0 & 91.1 & 95.2 & 95.1 \\
     & 2 & 88.8 & 88.2 & 82.2 & 91.7 & 94.9 & 94.9 \\
     & 3 & 73.8 & 73.1 & 95.1 & 94.4 & 90.4 & 90.8 \\
     & 4 & 65.9 & 65.0 & 89.1 & 94.0 & 94.7 & 94.7 \\
     & 5 & 64.2 & 63.6 & 83.6 & 91.7 & 95.2 & 95.0 \\
     & 6 & 65.7 & 64.6 & 85.1 & 92.7 & 94.9 & 94.9 \\
     \midrule
     Run time & & 1.55 & 3.17 & 3.00 & 6.17 & 3.31 & 5.24 \\
    \bottomrule
  \end{tabular}
  \end{center}
\end{table}

\newpage 

\section*{Figure Legends}

\hspace{5mm} Figure~\ref{fig:richFigure}: Predicted mean species richness for the eastern forest bird community (A) and the grassland bird community (C), as well as their associated standard deviations (B, D) using a spatial latent factor multi-species occupancy model (\texttt{sfMsPGOcc}).

\hspace{5mm} Figure~\ref{fig:richDiffFigure}: Difference in predicted mean richness from a spatial latent factor multispecies occupancy model to two simpler candidate models. Panels (A) and (B) show differences with the non-spatial latent factor multi-species occupancy model for the eastern forest and grassland bird communities, respectively, while panels (C) and (D) show differences with the spatial factor joint species distribution model.

\newpage

\section*{Figures}

\newpage

\newpage

\renewcommand{\thefigure}{S\arabic{figure}}
\renewcommand{\thetable}{S\arabic{table}}
\setcounter{figure}{0}
\setcounter{table}{0}

\noindent Doser, Jeffrey W., Finley, Andrew O., and Banerjee, Sudipto. Joint species distribution models with imperfect detection for high-dimensional spatial data. Submitted to Ecology.

\section*{Appendix S1}

\section*{Section S1 Candidate models}

To assess the benefits of accounting for residual species correlations, spatial autocorrelation, and imperfect detection, we compare the spatial factor multi-species occupancy model to five candidate models that only address a subset of the three complexities (Appendix S1: Table S1). We provide functionality for all five candidate models in the \texttt{spOccupancy} R package, and subsequently refer to all models by their \texttt{spOccupancy} function name (Appendix S1: Table S1). We use a \pg data augmentation approach in all models to yield computationally efficient Gibbs samplers (Appendix S2).

\begin{table}[ht!] % <--
  \begin{center}
  \caption{Characteristics of the six candidate models used in the simulation study and case study, as well as the function name for model implementation in the \texttt{spOccupancy} R package \citep{doser2022spoccupancy}.}
  \label{tab:candidateModels}
  \begin{tabular}{c c c c}
    \toprule
    \texttt{spOccupancy} & Species & Spatial & Imperfect \\
    Function & Correlations & Autocorrelation & Detection \\
    \midrule
    \texttt{lfJSDM} & \checkmark & & \\
    \texttt{sfJSDM} & \checkmark & \checkmark & \\
    \texttt{msPGOcc} & & & \checkmark \\
    \texttt{spMsPGOcc} & & \checkmark & \checkmark \\
    \texttt{lfMsPGOcc} & \checkmark & & \checkmark \\
    \texttt{sfMsPGOcc} & \checkmark & \checkmark  & \checkmark \\
    \bottomrule
  \end{tabular}
  \end{center}
\end{table}

\subsection*{Latent factor joint species distribution model (\texttt{lfJSDM})}

The latent factor JSDM (\texttt{lfJSDM}) is a standard joint species distribution model that ignores imperfect detection and spatial autocorrelation but accounts for species residual correlations. We account for species correlations using a latent factor model, where the latent factors arise from standard normal distributions instead of a spatial process as shown in Equation 4. This model is analogous to many varieties of non-spatial JSDMs that leverage a factor modeling approach for dimension reduction (e.g., \citealt{hui2016boral, ovaskainen2017make}). Because this model does not account for imperfect detection, we eliminate the detection sub-model and rather directly model a simplified version of the replicated detection-nondetection data, denoted as $y^*_i(\bm{s}_j)$, where $y^*_i(\bm{s}_j) = I(\sum_{k = 1}^{K_j}y_{i, k}(\bm{s}_j) > 0)$, with $I(\cdot)$ an indicator function denoting whether or not species $i$ was detected during at least one of the $K_j$ replicates at site $j$.

\subsection*{Spatial factor joint species distribution model (\texttt{sfJSDM})}

The spatial factor JSDM (\texttt{sfJSDM}) is a spatially-explicit joint species distribution model that ignores imperfect detection but accounts for spatial autocorrelation and species residual correlation. \texttt{sfJSDM} is analogous to \texttt{lfJSDM}, except the latent factors arise from a spatial process following Equation 4. This is the model presented by \cite{tikhonov2020computationally}.

\subsection*{Multi-species occupancy model (\texttt{msPGOcc})}

The multi-species occupancy model (\texttt{msPGOcc}) is a standard multi-species occupancy model \citep{dorazio2005} that accounts for imperfect detection but ignores spatial autocorrelation and species residual correlations. This model is identical to the full spatial factor multispecies occupancy model except the latent spatial process $\text{w}^*_i(\bm{s}_j)$ is removed from Equation 2. See \cite{doser2022spoccupancy} for additional details.

\subsection*{Spatial multi-species occupancy model (\texttt{spMsPGOcc})}

The spatial multi-species occupancy model (\texttt{spMsPGOcc}) extends \texttt{msPGOcc} by including a species-specific spatial process in the model for species-specific occurrence probability. Unlike \texttt{sfMsPGOcc}, here we assume each spatial process is independent of each other, resulting in a model where we need to estimate a spatial process for each species and subsequently ignore any residual species correlation. See \cite{doser2022spoccupancy} for additional details.

\subsection*{Latent factor multi-species occupancy model (\texttt{lfMsPGOcc})}

The latent factor multi-species occupancy model (\texttt{lfMsPGOcc}) is identical to \texttt{sfMsPGOcc} except we model the latent factors as standard normal random variables rather than from a spatial process as in Equation 4, and thus this model does not account for spatial autocorrelation. \texttt{lfMsPGOcc} is analogous to the latent variable model of \cite{tobler2019joint} except we use a logit link function and \pg latent variables rather than a probit formulation.

\section*{Section S2 Additional simulation study details}

For each data set, we simulated detection-nondetection data from $N = 10$ species at $J = 225$ sites with $K = 3$ replicates at each site. For all scenarios we assumed occurrence was a function of an intercept and 15 spatially-varying covariates. The intercept had a community-level mean of 0.2 and community-level variance of 1.5, which ultimately resulted in an average occurrence probability of $\sim0.55$ for a species in the community, with high variation in the species-specific occurrence probabilities. The community-level mean covariate effects were drawn from a uniform distribution with lower bound -1 and upper bound 1, and the community-level variances of the covariate effects were drawn from a uniform distribution with lower bound 0 and upper bound 2, resulting in low to high variation in the covariate effects across species in the community. For scenarios with imperfect detection (Scenarios 3-6), we assumed detection was a function of an intercept and five observation-level covariate effects. The intercept had a community-level mean of 0 and a community-level variance of 0.2, resulting in an average detection probability of 0.5 with moderate variation across the species-specific detection probabilities. We drew the detection community-level mean and variances of the covariate effects from the same uniform distributions as described for the occurrence level effects. For scenarios without imperfect detection (Scenarios 1-2), we assumed detection probability was constant with a value of $\pi = 0.8$, which corresponded to a probability of 0.992 for detecting the species during one of the three replicates if it was in fact present. While we could have assumed detection to be perfect (i.e., $\pi = 1$), this is highly unlikely in real ecological data, and rather a scenario of constant, high detection probability is more realistic and is often considered an adequate situation for fitting models that ignore imperfect detection. For Scenario 4, we assumed an exponential spatial correlation function and generated species-specific spatial range and spatial variance parameters from uniform distributions, while for Scenarios 2 and 6 we generated latent factor spatial processes from an exponential correlation function with spatial range parameters from a uniform distribution. We used $q = 3$ latent factors for all scenarios with species correlations (Scenarios 1, 2, 5, 6).

\section*{Section S3 Additional case study details}

Instead of using each of the 50 BBS stops as a spatial replicate in an occupancy modeling framework, we summarized the data into $K = 5$ replicates each comprised of data from ten BBS stops. In exploratory analyses, we found minimal differences between models using the full 50 stop data compared to our approach using the five replicates, as has been found in previous studies (e.g., \citealt{rushing2019modeling}). Further, using only five replicates results in substantial computational improvements as using all 50 BBS stops as replicates would increase the computational burden by needing to simulate more \pg random variables at each MCMC sample. Given the minimal differences between estimates using the full 50 stops, we used the five spatial replicates to minimize computational run times.

We modeled route-level occurrence of the 98 species as a function of five bioclimatic variables and eight land cover variables. We used five `bioclim` variables that are uncorrelated and perform well for modeling species distributions \citep{barbet201440, rushing2019modeling}: mean annual temperature, mean diurnal temperature range, mean temperature of the wettest quarter, total annual precipitation, and total precipitation of the warmest quarter. Following \cite{rushing2019modeling} and \cite{clement2016estimating}, we calculated the variables from the 12 months prior to the beginning of the BBS surveys (i.e., June 2017 - May 2018). We derived these variables from the Parameter-elevation Regression on Independent Slopes Model (PRISM; \citealt{daly2008physiographically}) project. PRISM provides monthly, high-resolution (4km) gridded data products on minimum/maximum temperatures and precipitation across the United States. We extracted the monthly values at the starting location of each BBS route for calculation of the five bioclimatic variables. We extracted the PRISM data in R \citep{rSoftware} using the \texttt{prism} \citep{prism} package and calculated the `bioclim` variables using the \texttt{dismo} package \cite{dismo}. We obtained land cover variables in 2018 from the USGS EROS (Earth Resources Observation and Science) Center, which produces high-resolution (250m) annual LULC maps across the continental US that are backcasted to 1938. For 2018, we calculated the proportion of water, barren land, forest, grassland, shrubland, hay, wetland, and developed land within a 5km radius circle centered around the starting location of each route.

To directly compare model predictive performance using models that do and do not account for imperfect detection, we fit all models using 75\% of the data points and kept the remaining 25\% of the data points for evaluation of model predictive performance. We collapsed the five spatial replicates at each site in the hold-out data set into a single value of 1 if the species was detected and 0 if not. We then compared predictions from each model to the collapsed data at the hold-out locations, and used the model deviance as a scoring rule of predictive performance \citep{hooten2015guide}, where lower values indicate better model predictive performance. Additionally, we compared predictions of latent occurrence at the hold out locations from each model to estimates of the latent occurrence state ($z_i(\bm{s}_j)$) generated from the three models that account for imperfect detection (\texttt{msPGOcc}, \texttt{lfMsPGOcc}, \texttt{sfMsPGOcc}) using the complete data set \citep{zipkin2012evaluating}. We then averaged across the three model deviance scoring rules to generate a single measure of predictive performance for the latent occurrence state. This allowed us to assess performance of the models in predicting the ecological process of interest rather than the raw detection-nondetection values (which confounds imperfect detection and true species occurrence) while accounting for model uncertainty \citep{doser2022integrated}.

\section*{Section S4 additional results and discussion}

\subsubsection*{Simulation study}

The spatial factor multi-species occupancy model (\texttt{sfMsPGOcc}) showed negligible differences in both bias and coverage rates compared to \texttt{spMsPGOcc} when data were simulated with an independent spatial process for each species (Scenario 4). Further, \texttt{sfMsPGOcc} had better coverage rates compared to \texttt{spMsPGOcc} when data were generated with species correlations and no spatial autocorrelation. Together with substantial decreases in run time for \texttt{sfMsPGOcc} compared to \texttt{spMsPGOcc}, this suggests \texttt{sfMsPGOcc} is a more efficient alternative to address spatial autocorrelation in multi-species detection-nondetection data sets.

\doublespacing

\begin{table}[ht!] % <--
  \begin{center}
  \caption{Estimated root mean squared error of simulated species-specific occurrence probabilities for six different simulation scenarios and six models of varying complexity. Bias values are averaged across all 10 species across 100 simulated data sets. Boldface indicates the model with the lowest bias for each simulation scenario. Scenario 1: residual species correlations and constant, high detection. Scenario 2: residual species correlations, constant and high detection, spatial autocorrelation. Scenario 3: imperfect detection only. Scenario 4: imperfect detection and spatial autocorrelation. Scenario 5: residual species correlations and imperfect detection. Scenario 6: residual species correlations, imperfect detection, and spatial autocorrelation}
  \label{tab:psiBias}
  \begin{tabular}{c c c c c c c}
    \toprule
    Scenario & & & Model & & & \\
    \midrule
     & \texttt{lfJSDM} & \texttt{sfJSDM} & \texttt{msPGOcc} & \texttt{spMsPGOcc} & \texttt{lfMsPGOcc} & \texttt{sfMsPGOcc} \\
     1 & 0.177 & 0.177 & 0.191 & 0.181 & \textbf{0.173} & 0.173 \\
     2 & 0.176 & 0.175 & 0.191 & 0.179 & 0.173 & \textbf{0.171} \\
     3 & 0.195 & 0.194 & \textbf{0.133} & 0.136 & 0.143 & 0.142 \\
     4 & 0.218 & 0.217 & 0.179 & \textbf{0.174} & 0.177 & 0.176 \\
     5 & 0.226 & 0.227 & 0.202 & 0.194 & \textbf{0.187} & 0.187 \\
     6 & 0.225 & 0.224 & 0.201 & 0.192 & 0.187 & \textbf{0.185} \\
    \bottomrule
  \end{tabular}
  \end{center}
\end{table}

\newpage

\begin{table}[ht!] % <--
  \begin{center}
  \caption{Estimated root mean squared error of simulated species-specific occurrence covariate effects for six different simulation scenarios and six models of varying complexity. Bias values are averaged across all 10 species and 15 estimated covariate effects across 100 simulated data sets. Boldface indicates the model with the lowest bias for each simulation scenario. Scenario 1: residual species correlations and constant, high detection. Scenario 2: residual species correlations, constant and high detection, spatial autocorrelation. Scenario 3: imperfect detection only. Scenario 4: imperfect detection and spatial autocorrelation. Scenario 5: residual species correlations and imperfect detection. Scenario 6: residual species correlations, imperfect detection, and spatial autocorrelation.}
  \label{tab:betaBias}
  \begin{tabular}{c c c c c c c}
    \toprule
    Scenario & & & Model & & & \\
    \midrule
     & \texttt{lfJSDM} & \texttt{sfJSDM} & \texttt{msPGOcc} & \texttt{spMsPGOcc} & \texttt{lfMsPGOcc} & \texttt{sfMsPGOcc} \\
     1 & 0.415 & 0.418 & 0.441 & \textbf{0.395} & 0.398 & 0.396 \\
     2 & 0.414 & 0.415 & 0.440 & \textbf{0.394} & 0.403 & 0.400 \\
     3 & 0.543 & 0.548 & \textbf{0.406} & 0.490 & 0.658 & 0.646 \\
     4 & 0.597 & 0.601 & 0.421 & \textbf{0.411} & 0.465 & 0.460 \\
     5 & 0.616 & 0.619 & 0.461 & \textbf{0.425} & 0.438 & 0.436 \\
     6 & 0.609 & 0.613 & 0.452 & \textbf{0.422} & 0.451 & 0.446 \\
    \bottomrule
  \end{tabular}
  \end{center}
\end{table}

\newpage

\subsection*{Case study}

The spatial factor multi-species occupancy model clearly distinguished between the two bird communities via the species-specific factor loadings and the latent spatial factors (Appendix S1: Figures S1-S5). In particular, a map of the first spatial factor across the US revealed high values in the eastern US (Appendix S1: Figure S1). Accordingly, 99\% of the mean species-specific factor loadings for the first spatial factor were greater than zero for the eastern forest bird community, compared to only 34\% of the factor loadings for the grassland bird community, indicating the potential of the spatial factor multi-species occupancy model to serve as a model-based ordination technique \citep{hui2015model}. The second spatial factor showed high values in the Great Plains region and to a lesser extent distinguished between the two communities, with 46\% and 91\% positive factor loadings for the eastern forest and grassland communities, respectively (Appendix S1: Figure S2). The additional three factors further distinguished between subsets of species within each community (Appendix S1: Figures S3-S5). As an alternative to interpreting the spatial factors and species-specific factor loadings, we can recover a full species-to-species covariance matrix using the factor loadings matrix as $\bm{\Lambda}\bm{\Lambda}^{\top}$, which, while singular, may be able to provide insight on the residual co-occurrence patterns between pairs of species in the modeled community. This can be the result of missing environmental drivers, biological interactions, and/or model mis-specification. While the covariance matrix provides information on which species tend to occur together, we caution against interpretation of these covariances as true biological interactions, as co-occurrence does not imply an interaction \citep{poggiato2021interpretations}.

We found slow MCMC convergence and mixing of the species-specific factor loadings in the spatial factor multi-species occupancy model for communities of species with a large number of rare species. This is in large part due to weak identifiability of the factor loadings ($\bm{\lambda}_i(\bm{s}_j)$) and spatial factors ($\textbf{w}(\bm{s}_j)$), as it is only their product ($\bm{\lambda}_i(\bm{s}_j)^\top\textbf{w}(\bm{s}_j)$) that influences species-specific occurrence probability. Further, the identifiability constraints placed on $\bm{\Lambda}$ requires consideration of the first $q$ species in the detection-nondetection data array, as certain factor loadings are fixed for these species. See Appendices S3 and S4 for further discussion of these challenges and how to address them when fitting models in \texttt{spOccupancy}.

\begin{table}[ht!] % <--
  \begin{center}
  \caption{Out-of-sample model deviance and WAIC results for the five candidate models used in the breeding bird case study. Boldface indicates the best performing model according to the given criteria. Note that WAIC is only comparable within models that do or do not account for imperfect detection. \texttt{spMsPGOcc} was not considered due to the massive computation time required.}
  \label{tab:modelSelection}
  \begin{tabular}{c c c c c c}
    \toprule
    Deviance type & & & Model & & \\
    \midrule
     & \texttt{lfJSDM} & \texttt{sfJSDM} & \texttt{msPGOcc} & \texttt{lfMsPGOcc} & \texttt{sfMsPGOcc} \\
     Data & 344 & \textbf{233} & 398 & 431 & 327 \\
     Latent & 411 & 293 & 385 & 372 & \textbf{243} \\
     \midrule
     WAIC & 87,615 & \textbf{84,192} & 417,954 & 395,094 & \textbf{390,607}\\
    \bottomrule
  \end{tabular}
  \end{center}
\end{table}

\newpage

\begin{figure}[ht]
    \centering
    \includegraphics[width=15cm]{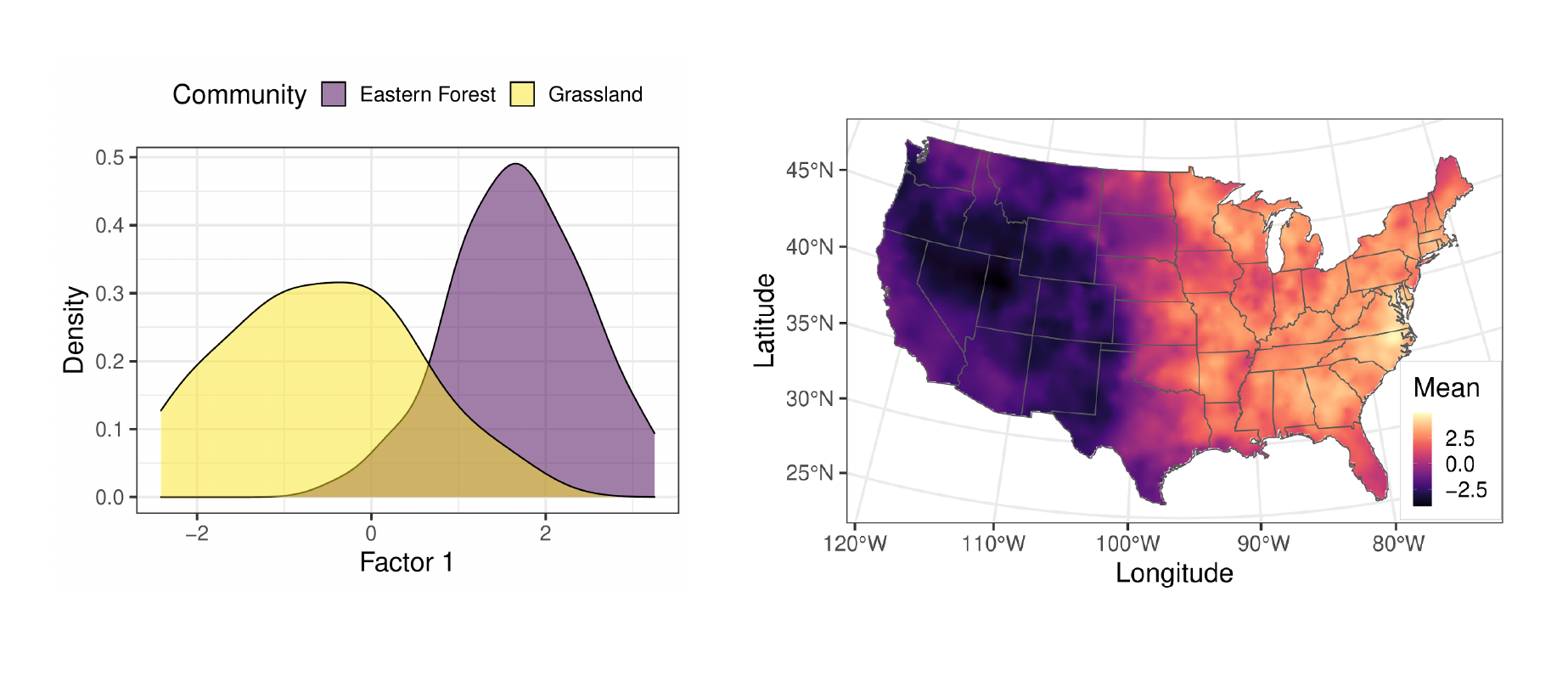}
    \caption{Density of estimated mean species-specific factor loadings for all species in the eastern forest and grassland bird communities on the first latent spatial factor, and the associated mean realization of the spatial factor across the US.}
    \label{fig:factor1}
\end{figure}

\newpage

\begin{figure}[ht]
    \centering
    \includegraphics[width=15cm]{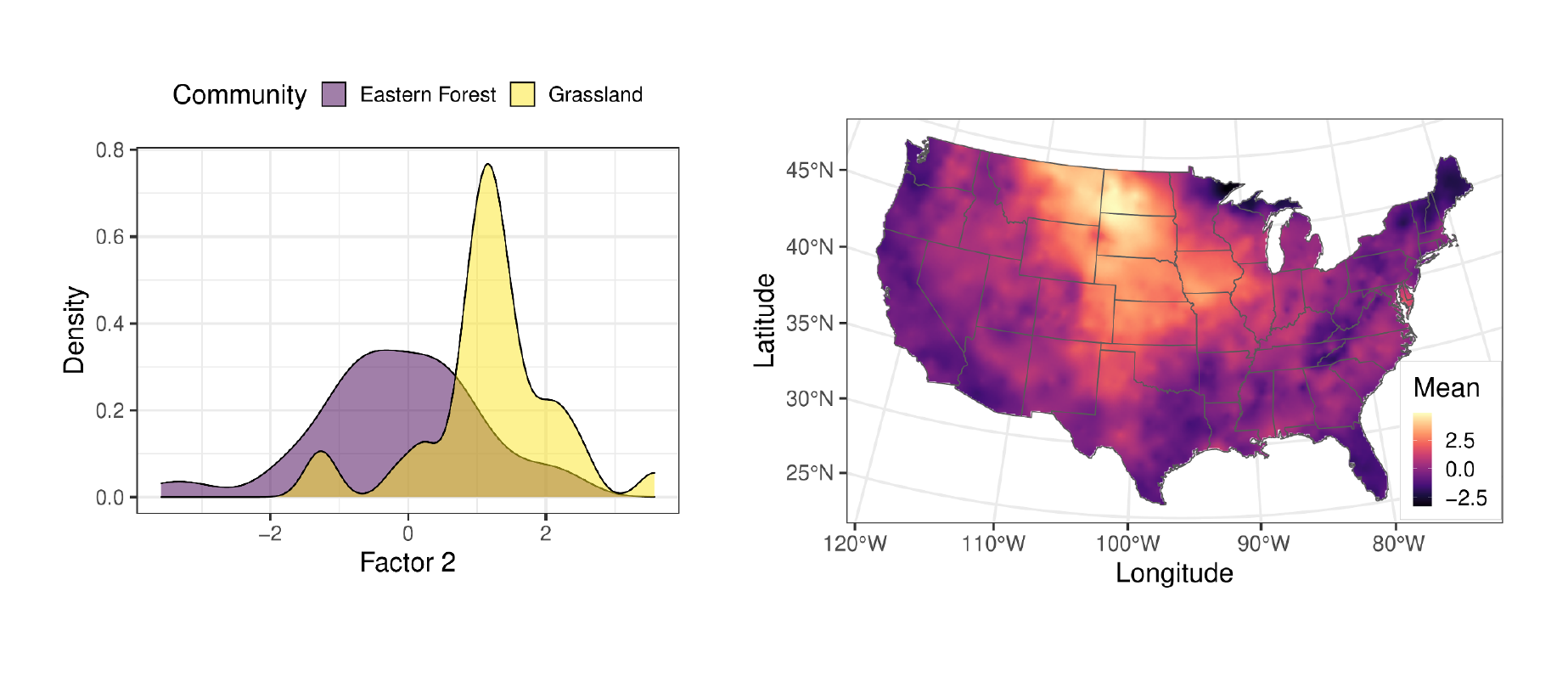}
    \caption{Density of estimated mean species-specific factor loadings for all species in the eastern forest and grassland bird communities on the second latent spatial factor, and the associated mean realization of the spatial factor across the US.}
    \label{fig:factor2}
\end{figure}

\newpage

\begin{figure}[ht]
    \centering
    \includegraphics[width=15cm]{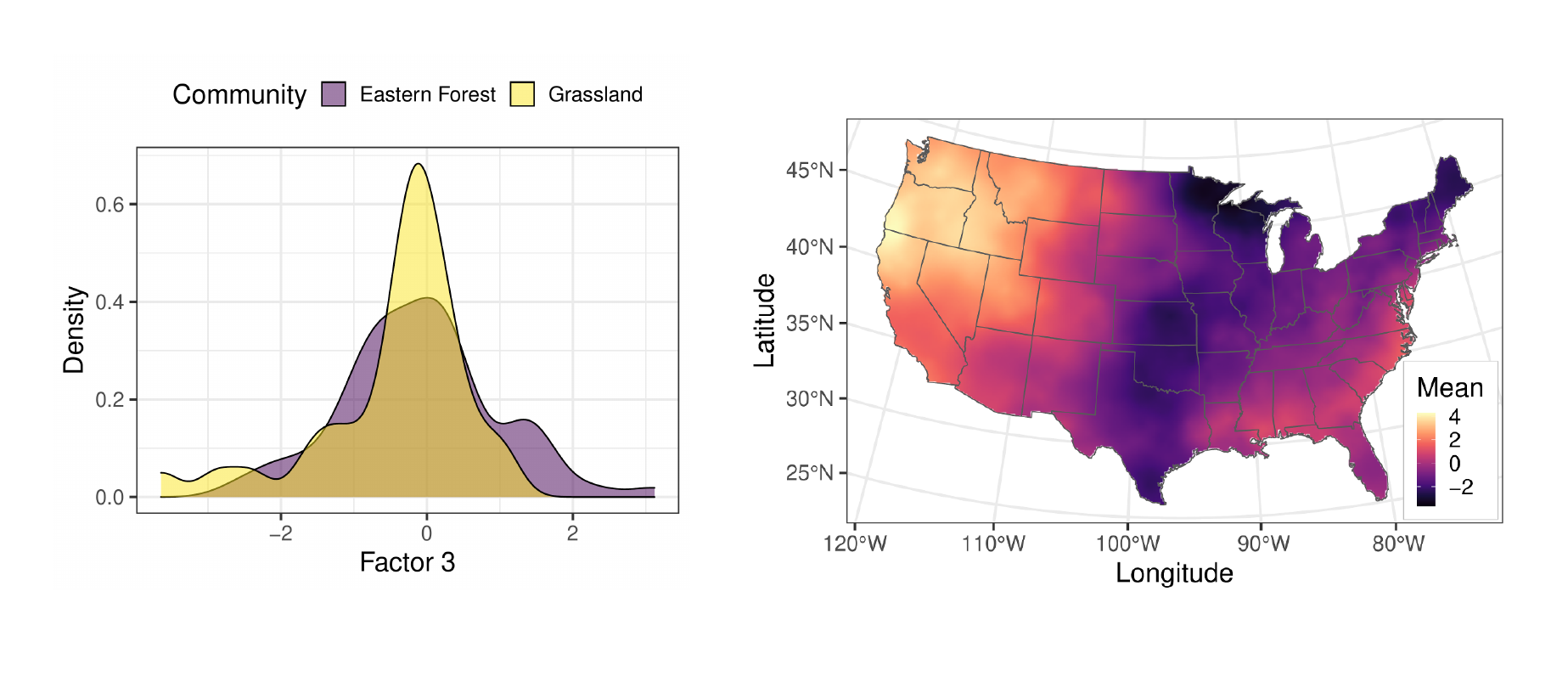}
    \caption{Density of estimated mean species-specific factor loadings for all species in the eastern forest and grassland bird communities on the third latent spatial factor, and the associated mean realization of the spatial factor across the US.}
    \label{fig:factor3}
\end{figure}

\newpage

\begin{figure}[ht]
    \centering
    \includegraphics[width=15cm]{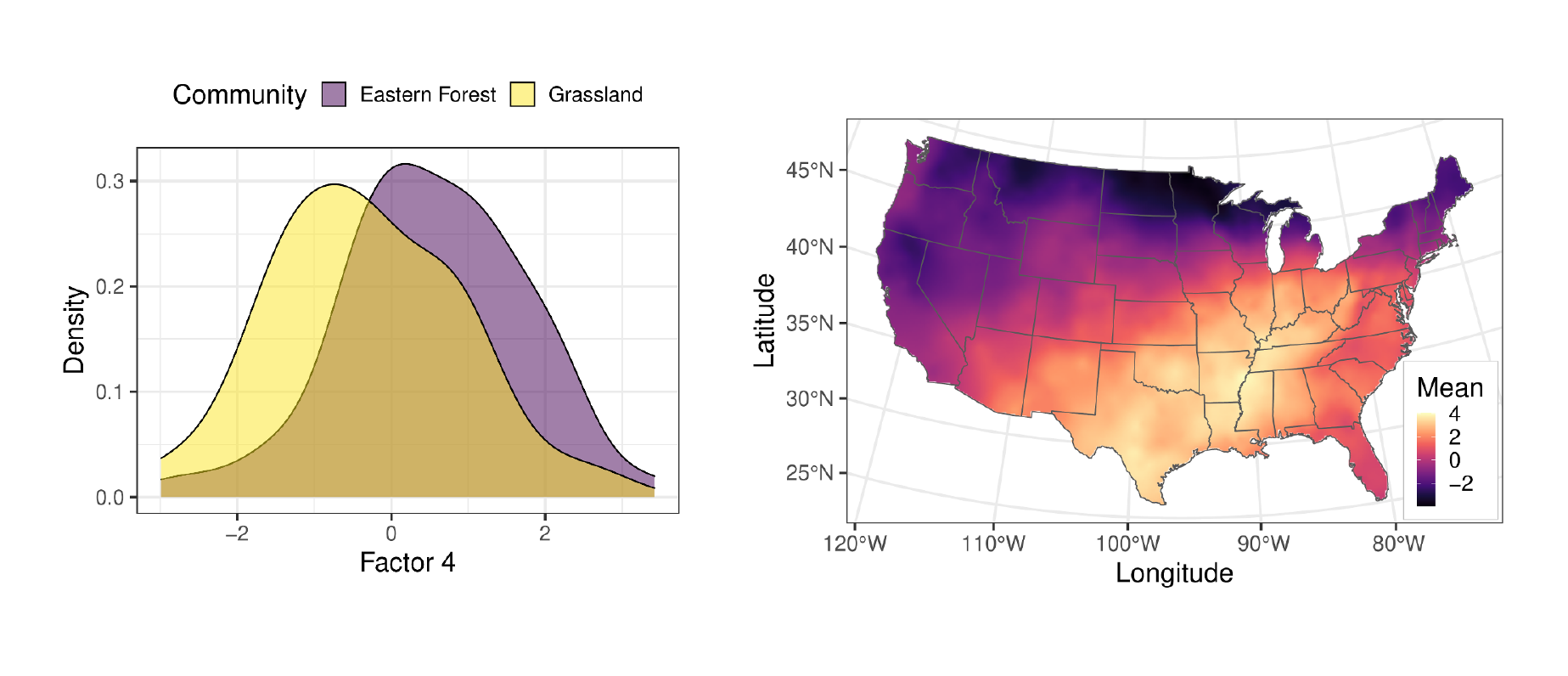}
    \caption{Density of estimated mean species-specific factor loadings for all species in the eastern forest and grassland bird communities on the fourth latent spatial factor, and the associated mean realization of the spatial factor across the US.}
    \label{fig:factor4}
\end{figure}

\newpage

\begin{figure}[ht]
    \centering
    \includegraphics[width=15cm]{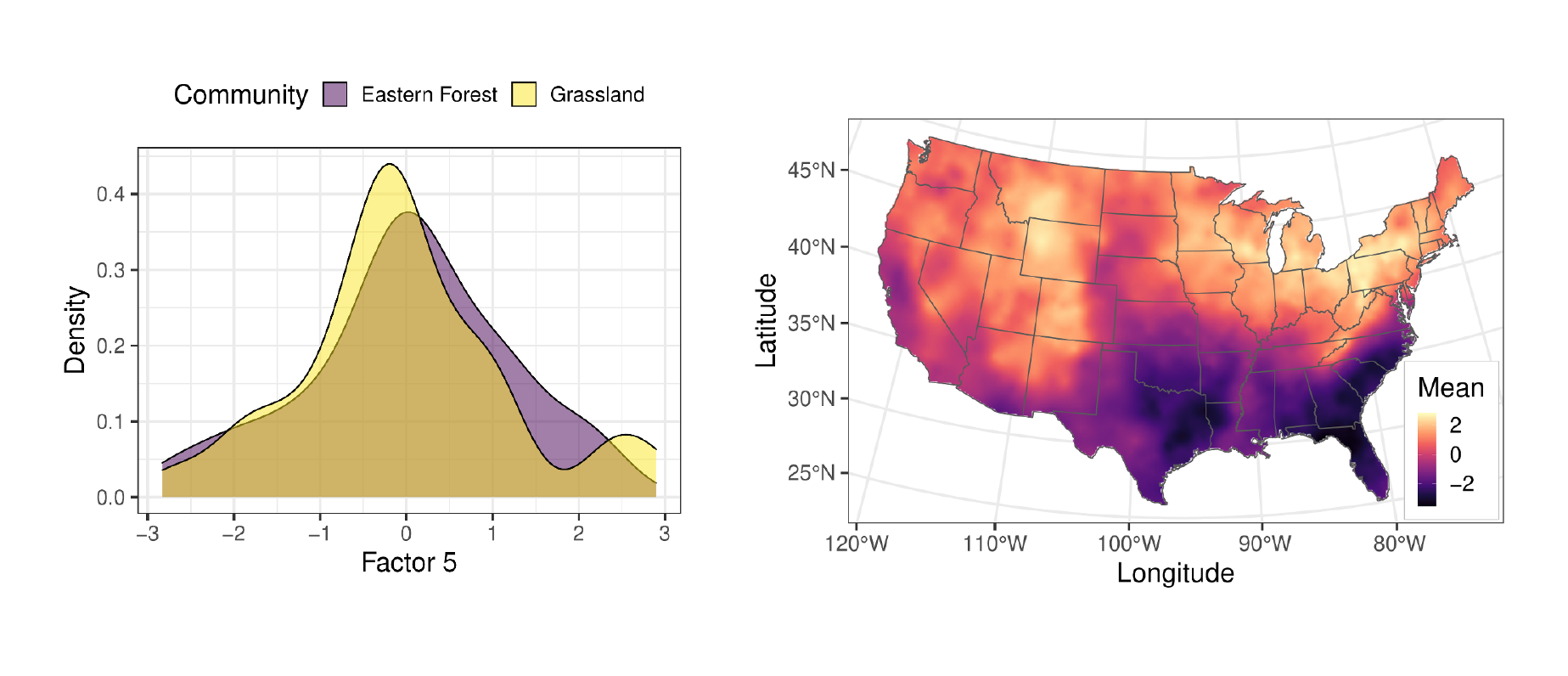}
    \caption{Density of estimated mean species-specific factor loadings for all species in the eastern forest and grassland bird communities on the fifth latent spatial factor, and the associated mean realization of the spatial factor across the US.}
    \label{fig:factor5}
\end{figure}

\newpage

\noindent Doser, Jeffrey W., Finley, Andrew O., and Banerjee, Sudipto. Joint species distribution models with imperfect detection for high-dimensional spatial data. Submitted to Ecology.

\section*{Appendix S4: Determining the number of factors in a spatial factor multi-species occupancy model.}

\doublespacing

Determining the number of latent factors to include in a spatial factor multi-species occupancy model or non-spatial latent factor multi-species occupancy model is not straightforward. Often using as few as 2-5 factors is sufficient, but for particularly large communities (e.g., $N = 600$), a larger number of factors may be necessary to accurately represent variability among the species \citep{tobler2019joint, tikhonov2020computationally}. While other approaches exist to estimate the ``optimal'' number of factors directly in the modeling framework \citep{tikhonov2020computationally, ovaskainen2016uncovering}, these approaches do not allow for interpretability of the latent factors and the latent factor loadings (see Appendix S1: Figures S1-S5). The specific restraints and priors we place on the factor loadings matrix ($\bm{\Lambda}$) in our approach allows for interpretation of the latent factors and the factor loadings, but does not automatically determine the number of factors for optimal predictive performance. Thus, there is a trade-off between interpretability of the latent factors and factor loadings and optimal predictive performance. In our \texttt{spOccupancy} implementation, we chose to allow for interpretability of the factor and factor loadings at risk of inferior predictive performance if too many or too few factors are specified by the user.

The number of latent factors can range from 1 to $N$ (the total number of species in the modeled community). Conceptually, choosing the number of factors is similar to performing a principal components analysis and looking at which components explain a large amount of variation. We want to choose the number of factors that explains an adequate amount of variability among species in the community, but we want to keep this number as small as possible to avoid overfitting the model and large model run times. When initially specifying the number of factors, we suggest the following:

\begin{enumerate}
    \item Consider the size of the community and how much variation there is between species. If there is expected large variation in occurrence patterns for all species in the community, the user may require a larger number of factors. If the modeled community is comprised of certain groups of species that are expected to behave similarly (e.g., insectivores, frugivores, granivores), then a smaller number of factors may suffice. Further, as shown by \cite{tikhonov2020computationally}, as the number of species increases, more factors will likely be necessary to adequately represent variability in the community.
    \item Consider the amount of computational time/power that is available. Model run time to achieve convergence will increase as more factors are included in the model. Under certain circumstances (i.e., there is a large number of spatial locations in the data set), reasonal run times may only be possible with a modest or small number of factors.
    \item Consider the rarity of species in the community, how many spatial locations (i.e., sites) are in the data set, and how many replicates are available at each site. Models with more latent factors have more parameters to estimate, and thus require more data. If there are large numbers of rare species in the community (like in the BBS case study in the main text), we may be limited in the number of factors we can specify in the model, as models with more than a few factors may not be identifiable. Such a problem may also arise when working with a small number of spatial locations (e.g., 30 sites) or replicates (e.g., 1 or 2 replicates at each site).
\end{enumerate}

Because of the restrictions we place on the factor loadings matrix $\bm{\Lambda}$ (diagonal elements equal to 1 and upper triangle elements equal to 0), the user must also carefully consider the order of species in the detection-nondetection data array. More specifically, we need to choose the first $q$ species in the array (where $q$ is the number of latent factors in the model), as these are the species that will have restrictions on their factor loadings. While from a theoretical perspective the order of the species will only influence the resulting interpretation of the latent factors and factor loadings matrix and not the model estimates, this decision does have practical implications. We have found that careful consideration of the ordering of species can lead to (1) increased interpretability of the factors and factor loadings; (2) faster model convergence; and (3) improved mixing. Determining the order of the factors is less important when there are an adequate number of observations for all species in the community, but it becomes increasingly important as more rare species are present in the data set. If encountering difficulty when fitting a spatial/latent factor multi-species occupancy model in \texttt{spOccupancy} (e.g., MCMC chains are not converging or have extremely slow mixing), we suggest the following when considering the order of the species in the detection-nondetection array:

\begin{enumerate}
  \item Place a common species first. The first species has all of its factor loadings set to fixed values, and so it can have a large influence on the resulting interpretation of the factor loadings and latent factors. We have also found that having a rare species first can result in slow mixing of the MCMC chains and increased sensitivity to initial values of the latent factor loadings matrix.
  \item For the remaining $q - 1$ factors, place species that are \textit{a priori} believed to show different occurrence patterns than the first species, as well as the other species placed before it. Place these remaining $q - 1$ species in order of decreasing differences from the initial factor. For example, if we fit a spatial factor multi-species occupancy model with three latent factors ($q = 3$) and were encountering difficult convergence of the MCMC chains, for the second species in the array, we would place a species that we believed shows large differences in occurrence patterns from the first species. For the third species in the array, we would place a species that we believed to show different occurrence patterns than both the first and second species, but its patterns may not be as noticeably different compared to the differences between the first and second species.
\end{enumerate}

After successfully fitting a spatial/latent factor multi-species occupancy model in \texttt{spOccupancy}, we can look at the posterior summaries of the latent factor loadings to provide information on how many factors are necessary for the given data set. In particular, we can look at the posterior mean or median of the latent factor loadings for each factor. If the factor loadings for all species are very close to zero for a given factor, that suggests that specific factor is not an important driver of species-specific occurrence across space, and thus we may consider removing it from the model. Additionally, we can look at the 95\% credible intervals, and if the 95\% credible intervals for the factor loadings of all species for a specific factor all contain zero this is further support to reduce the number of factors in the model.

\end{document}